\documentclass[%
 aip,
 amsmath,amssymb,
 reprint,%
]{revtex4-2}

\usepackage{graphicx}
\usepackage{color}
\usepackage{amsfonts}
\usepackage{amsmath}
\usepackage{textcomp}
\usepackage{setspace}

\begin{document}

\title{Micro- and nanoscale focusing across the XUV range of the ASTRID2 light source with a capillary optic}
\author{Alfred J. H. Jones}
\email{ajones@phys.au.dk}
\affiliation{Department of Physics and Astronomy, Aarhus University, 8000 Aarhus C, Denmark}
\author{Zhihao Jiang}
\affiliation{Department of Physics and Astronomy, Aarhus University, 8000 Aarhus C, Denmark}
\author{Asger Petersen}
\affiliation{Department of Physics and Astronomy, Aarhus University, 8000 Aarhus C, Denmark}
\author{Søren V. Hoffmann}
\affiliation{Department of Physics and Astronomy, Aarhus University, 8000 Aarhus C, Denmark}
\author{Nykola C. Jones}
\affiliation{Department of Physics and Astronomy, Aarhus University, 8000 Aarhus C, Denmark}
\author{Philip Hofmann}
\affiliation{Department of Physics and Astronomy, Aarhus University, 8000 Aarhus C, Denmark}
\author{Jill A. Miwa}
\affiliation{Department of Physics and Astronomy, Aarhus University, 8000 Aarhus C, Denmark}
\author{Søren Ulstrup}
\affiliation{Department of Physics and Astronomy, Aarhus University, 8000 Aarhus C, Denmark}

\date{\today}

\begin{abstract}

Focusing of synchrotron light across extreme ultraviolet (XUV) and soft X-ray regimes is increasingly desired for photoemission-based techniques where reduced beam width gives access to smaller samples such as microscopic single crystals and functioning two-dimensional (2D) heterostructures and devices. Many existing focusing methods, however, are not able to take full advantage of the synchrotron beam due to limited photon energy range or low transmission. Modern capillary optics have enabled achromatic, high transmission focusing of XUV and X-ray light. Here, we present a detailed characterisation of such an achromatic capillary optic installed at the AU-SGM4 beamline for spatial- and angle-resolved photoemission spectroscopy  (ARPES) experiments at the ASTRID2 light source. The transmission of the capillary as a function of photon energy is given, and the dependence of the beam width, position, and transmission are measured against the source size. Analysis of the far-field image of the beam allows for slope errors on the inner surface of the capillary to be overcome by selectively aperturing the beam, resulting in a minimum beam width of 900 nm measured in a photoemission geometry. 

\end{abstract}

\maketitle

\section{Introduction}

Angle-resolved photoemission spectroscopy (ARPES) is an experimental technique capable of extracting the spectral function from crystalline solids. Traditionally requiring clean, uniform sample surfaces of the order of 100\,\textmu m, ARPES is increasingly targeting new materials hosting smaller, inhomogeneous surfaces \cite{Mo2017,Iwasawa2020}. These include artificially stacked heterostructures of layered 2D materials \cite{Avila2013,Gomez2014,Novoselov2016,Zhang2018} and operational 2D devices \cite{Nguyen2019,Hofmann2021}, where the relevant region of the structure is often less than 10\,\textmu m across, or in-vacuum cleaved crystals with inhomogeneous surfaces or multiple surface terminations \cite{Iwasawa2019,Watson2019,Volckaert2023}. 
Small areas (\textless10\,\textmu m) on a sample surface may be isolated either by uniformly illuminating the surface and then selectively measuring photoelectrons originating from the small region, or by focusing the XUV light to excite photoelectrons only from the desired area, which may then be measured using conventional ARPES detection techniques \cite{Maklar2020}. The former method typically utilizes strong electric fields in a photoemission electron microscopy (PEEM) lens followed by an energy filter, with a field aperture inside the lens tube, for selecting the area to measure \cite{Escher2005,Cattelan2018}. The latter method of focusing light to a small spot allows for the large electric fields to be avoided, enabling a greater range of samples for measurement. A smaller area may also be isolated for measurement using a focused XUV beam. However, use of a small spot puts stringent demands for precision manipulation of both the optic and sample. Additionally, space charge at the focal point may become a limiting factor for intense light sources with very small foci \cite{Bostwick2012,Rotenberg2014}.

A powerful tool for synchrotron-based ARPES is to vary the incident photon energy to gain access to the out-of-plane electronic dispersion of the target material. Fresnel zone plates, used to achieve beam spots of the order of 200\,nm \cite{Kastl2019,Rosner2019,Jones2022} in photoemission experiments, only focus the coherent fraction of the source, and cannot provide a constant spot with varying photon energy. In contrast, pairs of Kirkpatrick–Baez (KB) mirrors provide an achromatic transmission, but they only have been used to reach $\sim$10\,\textmu m beam widths in ARPES measurements \cite{Gong2016,Kitamura2022,Polley2024}. Recently, achromatic X-ray capillary optics have been implemented for photoemission measurements in the XUV and soft X-ray regimes at several synchrotron beamlines \cite{Koch2018,Avila2023,Nunn2023} to bridge the gap between these different focusing techniques. Capillary optics combine high transmission across a broad photon energy range with spot sizes less than 5\,\textmu m. Hence, the capillary optic is an excellent source for target cleaved and 2D heterostructure samples.

\begin{figure*} [ht!]
	\begin{center}
		\includegraphics[width=1\textwidth]{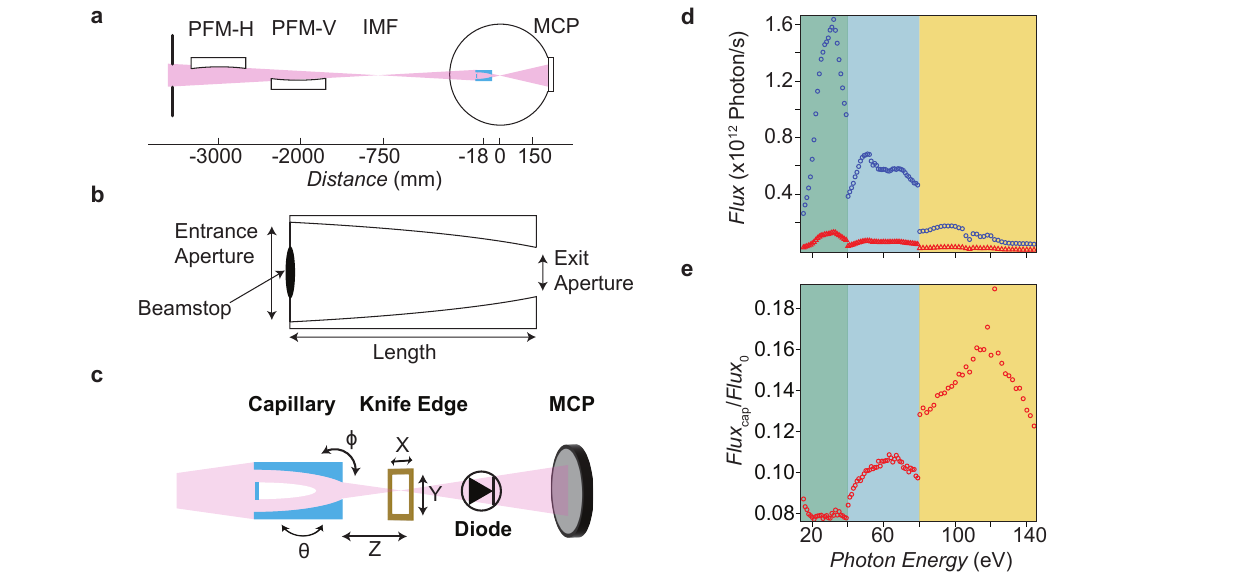}
		\caption{X-ray capillary optic at AU-SGM4. \textbf{a,} Schematic of the optical elements of the beamline following the monochromator exit slit. Scrapers used to aperture the beam are located before the PFM-H. All positions are referenced to the centre of the measurement chamber. \textbf{b,} Labeled diagram of the key components of the capillary at AU-SGM4, not to scale. \textbf{c,} Schematic of the focusing geometry used for focusing the capillary with the control parameters labeled. The location of the silicon photodiode used for flux measurements is marked. \textbf{d,} Photon flux measured on a silicon photodiode without the capillary (blue circles) and with the capillary in place (red triangles) for different photon energies. \textbf{e,} Transmission of the capillary as a function of photon energy. The coloured sections in \textbf{d} and \textbf{e} denote the different monochromator gratings: LEG = Green, MEG = Blue, HEG = Yellow \cite{Jones2025}.}
		\label{fig:Figure1}
	\end{center}
\end{figure*}

In this work, we describe in detail the performance of an X-ray capillary optic installed at the AU-SGM4 beamline of the ASTRID2 synchrotron, Aarhus, Denmark \cite{Hertel2011,Jones2025}. First, the focusing geometry utilized at the beamline is described, and the transmission of the capillary determined. Following this, a detailed characterisation of the beam close to its focus is made. Finally, using selective aperturing of the beam, a sub-micrometer beam is produced and the implications for spatially-resolved photoemission measurements are presented.

\section{Beamline Setup}
In this section, we describe the beam focusing geometry implemented at the AU-SGM4 beamline at ASTRID2 in order to set the stage for the detailed capillary characterization that follows. At AU-SGM4, light is taken from an undulator insertion device on ASTRID2 and monochromatised by a single-reflection, spherical grating monochromator (SGM) which uses a movable exit slit. The diverging light from the exit slit is re-focused by a pair of cylindrical post-focusing mirrors (PFMs) as sketched in Figure \ref{fig:Figure1}a. These produce an intermediate focus (IMF) 750\,mm from the sample. An elliptical capillary optic with a length of 26\,mm then images the IMF. A beamstop at the capillary entrance blocks all direct transmission and selects how much of the capillary’s inside surface is used for focusing. The incident light makes a single reflection on the inner, platinum surface, towards a focus 7.8\,mm from the exit of the capillary optic. The beam then diverges until a micro channel plate (MCP) detector 150\,mm away measures the far-field of the beam focus. The layout of the capillary is sketched in Figure \ref{fig:Figure1}b with key parameters of the capillary optic and its internal elliptical profile summarized in Table \ref{fig:CapTable}. 

\begin{table}[]
	\begin{tabular}{lllll}
		Semi-major axis & 375 mm &  &  &  \\
		Semi-minor axis & 4.06 mm &  &  &  \\
		Length & 25.8 mm &  &  &  \\
		Working distance & 7.8 mm &  &  & \\
		Entrance aperture & 3.36 mm & & &\\
		Exit aperture & 1.64 mm
	\end{tabular}
	\caption{Key parameters defining the ellipse of the used capillary optic. A diagram describing these is presented in Figure \ref{fig:Figure1}.}
	\label{fig:CapTable}
\end{table}

The transmission of the capillary was quantified using an XUV silicon diode, AXUV100G from Opto Diode Corp, placed 20\,mm downstream of the beam focus to avoid saturation and beam damage of the diode. The synchrotron current was 180\,mA and the monochromator entrance and exit slits set to 50\,\textmu m and 30\,\textmu m, respectively. The photocurrent measured on the diode with and without the capillary inserted into the beam was converted to photon flux using data acquired on an identical, calibrated photodiode. Figure \ref{fig:Figure1}d shows the measured photon flux with the photodiode illuminated with the full synchrotron beam (blue) and with the light transmitted through the capillary (red). The measured current of the full beam is in line with previous measurements from the same undulator on the AU-SGM3 beamline of ASTRID2 \cite{Hoffmann2004}. Upon changing grating there is a drop in flux, for the same photon energy, due to the reduced line spacing of the higher energy gratings reducing the transmission. The dip in intensity at 108\,eV corresponds to changing of the undulator harmonic from 3rd to 5th. The flux transmitted through the capillary follows the same general trend, with similar features apparent. The transmission of the capillary, defined as the ratio of flux through the capillary I$_{cap}$ to the initial photon flux I$_0$ is presented in Figure \ref{fig:Figure1}e. These demonstrate a smoothly varying transmission across each grating, with a jump in transmission between different gratings. This jump may be due to changes in the pointing of the beam with changing grating, causing changes in the illumination of the capillary, which would lead to a change in the transmission. The overall measured transmission is between 8$\%$ and 16$\%$, with the losses arising from light blocked by the beamstop, over-illumination of the capillary, and the reflectivity of the platinum coating. These values are ~10\textendash100$\times$ larger than the transmission of typical zone plate optics on third generation synchrotron sources \cite{Rosner2019}. 

\section{Capillary Focussing}

\begin{figure*} [t!]
	\begin{center}
		\includegraphics[width=1\textwidth]{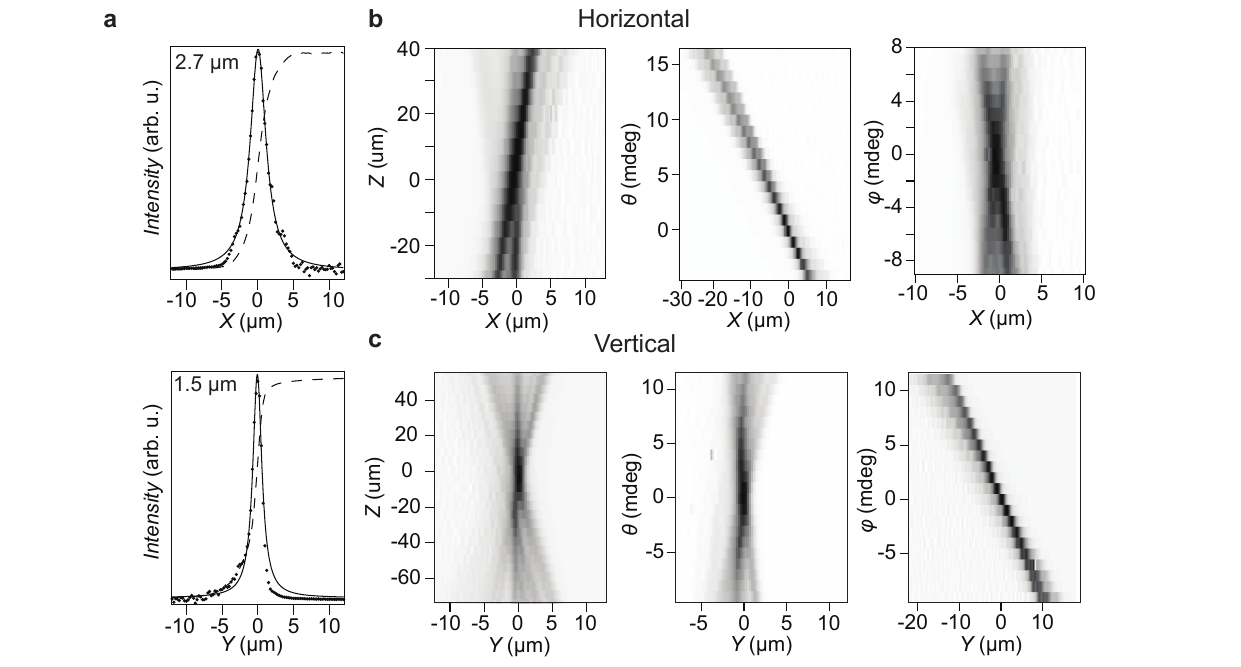}
		\caption{Focusing properties of X-ray capillary. \textbf{a,} Intensity profile of the beam at focus horizontally (top) and vertically (bottom). Measured intensity profile is marked with dashed lines, the differentiated intensity as markers, a Lorentzian fit to the differential intensity is a solid line. The FWHM is written in the upper left. \textbf{b,} Vertical differential profiles measured for different focal distance (\textit{Z}), horizontal capillary angle ($\theta$), and vertical capillary angle ($\phi$). \textbf{c,} Differential profiles of the beam measured horizontally.}
		\label{fig:Figure2}
	\end{center}
\end{figure*}

To characterise the focus generated by the capillary, a square opening with sharp, gold-coated knife edges is translated through the beam in the (\textit{X,Y})-plane perpendicular to the beam direction. The summed intensity measured on the MCP is recorded as a function of distance along the beam direction (\textit{Z}) and the vertical ($\phi$) or horizontal ($\theta$) angle of the capillary optic to the beam. The intensity of the beam as a function of the position in \textit{X} or \textit{Y} of the knife edge is then differentiated. The optimal profiles measured are demonstrated in Figure \ref{fig:Figure2}a with the measured intensity and differentiated intensity marked with dashed lines and dots, respectively. For a perfectly sharp knife edge, the differential intensity corresponds to the beam profile in that direction. Fitting to a Lorentzian (solid line), a minimum beam width of 2.7\,\textmu m horizontally, and 1.5\,\textmu m vertically is determined. The expected beam profile should be approximately Gaussian, however the longer tails in the data are better described by Lorentzian curves. We explore which parts of the focusing process primarily contribute to these profiles below.

Alignment of the capillary for measurement requires a high degree of precision and stability with beamline parameters. To determine sensitivity of the capillary alignment, the differential profile of the beam is measured for the main alignment parameters: The distance Z from the end of the capillary, and the angles $\phi$ and $\theta$ of the capillary. Figure \ref{fig:Figure2}b shows how the single peak splits into defined branches as \textit{Z} is varied away from the beam focus. Similarly, the beam width is highly sensitive to angle of the capillary optic, doubling the width with \textless10\,milidegree of misalignment in $\phi$. To determine the optimal focus shown in Figure \ref{fig:Figure2}a, the horizontal and vertical differential profiles are minimised with respect to $\phi$, $\theta$, \textit{Z}. Mapping a wide 3D (\textit{Z}, $\phi$, $\theta$) space ensures local minima of beam size are avoided, though this greatly increases the time required for focusing. The effect of moving the capillary horizontally and vertically is not found to improve the resulting beam focus, as shown in Figure \ref{fig:FigureA1} in Appendix \ref{app:capillary}. Instead, the focus shifts rigidly with the capillary, but also broadens, precluding this as a method for fine scanning on the sample.

\begin{figure*} [t!]
	\begin{center}
		\includegraphics[width=1\textwidth]{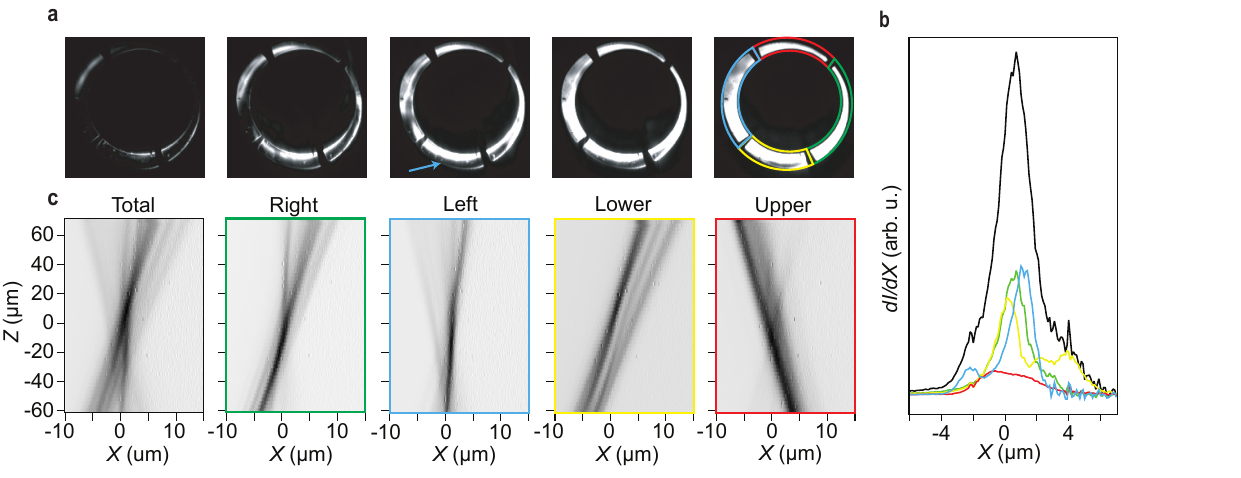}
		\caption{Detailed focusing behaviour. \textbf{a,} Far field intensity of the capillary as the knife edge is moved horizontally through the focus. \textbf{b,} Differential intensity profile (black line) and components from four quadrants of the far field image (coloured lines). The selected regions are marked in \textbf{a}. \textbf{c,} Variation in differential intensity profiles at positions along the focus axis for different, labeled, quadrants of the far field intensity.}
		\label{fig:Figure3}
	\end{center}
\end{figure*}

\begin{figure*} [t!]
	\begin{center}
		\includegraphics[width=1\textwidth]{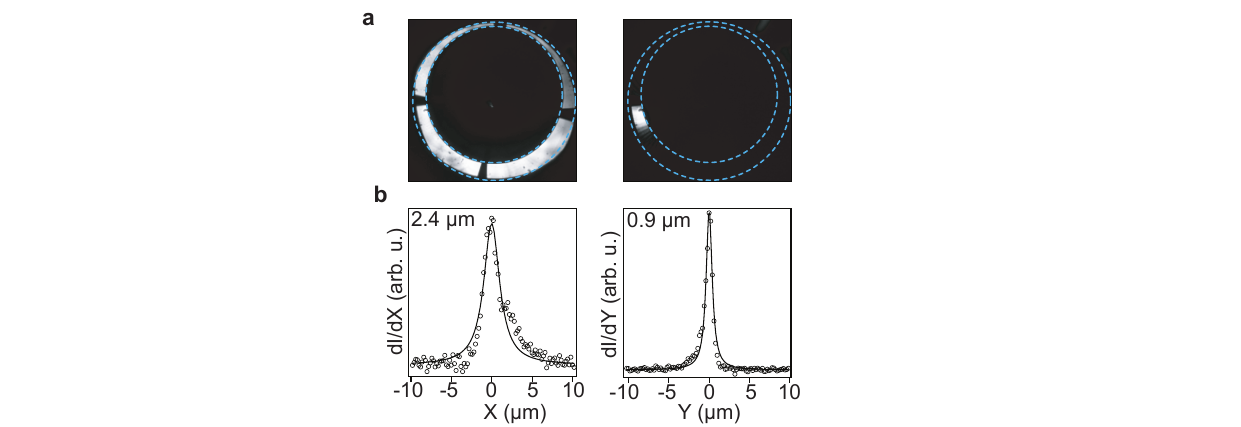}
		\caption{Selective aperturing of the beam for sub-micrometer spot sizes. \textbf{a,} Far-field images of the beam before (left) and after (right) aperturing the beam. \textbf{b,} Horizontal (left) and vertical (right) differential profiles of the apertured beam measured with the knife edge. The measured data are shown with markers, Lorentzian fits as solid lines.}
		\label{fig:Figure4}
	\end{center}
\end{figure*}

The measured beam focus of 2.7 $\times$ 1.5\,\textmu m is larger than the predicted value of ~1.5 $\times$1\,\textmu m, which is determined from the geometry of the beamline and capillary parameters, as discussed in ref.[\citenum{Jones2025}]. To determine the cause of this broadening, the far field image on the MCP was recorded for different positions of the knife edge cutting through the focus, with the capillary aligned such that the beam width was minimised. Figure \ref{fig:Figure3}a shows a series of images of the MCP, ranging from the MCP almost fully occluding the focus (left) to the knife edge being fully removed from the beam (right). While the change in intensity is generally uniform, indicating that the knife edge is placed at the optimal focal distance, the illumination is patchy, see for example the crescent in the lower region of the image, indicated with a blue arrow. To examine this non-uniformity in more detail, the image was split into four quadrants marked with coloured outlines in Figure \ref{fig:Figure3}a. The differential intensity profile for each of these quadrants in Figure \ref{fig:Figure3}b, effectively the beam profile originating from this portion of the capillary, show markedly different behaviors in each quadrant. The tails of overall differential intensity profile (black) are filled out by side peaks from different quadrants. The evolution of the beam profile away from the focus is shown in Figure \ref{fig:Figure3}c for the total intensity and for each of the quadrants. Different regions of the capillary clearly focus at different depths with no optimal position available where all are minimised simultaneously. The intensity distribution in the quadrants is not uniform. The quadrants were merely selected as convenient demonstrations of the overall behaviour. The results shown in Figure \ref{fig:Figure3} suggest that the broadened beam may be due to imperfections in the capillary, in particular slope errors in the inner, reflective surface.

The beam width may be reduced by selecting only one small, uniformly focusing region of the optic, at the cost of greatly reducing transmission. The beamstop at the entrance to the capillary may be used to isolate regions of the optic to this effect. Figure \ref{fig:FigureA2} in Appendix \ref{app:capillary} compares the focus achieved using two different beamstops, with the beamstop blocking more light producing a significantly reduced focus size.

To explore the possibility of further reducing the beam size, the beam was apertured using scrapers placed before the PFM-H mirror where the beam is largest, cutting the beam through the capillary down from the full illumination shown in the left-hand image of Figure \ref{fig:Figure4}a to the small square region in the right-hand image of Figure \ref{fig:Figure4}a. Knife-edge focusing was repeated on the beam from this selected area of the capillary, with the differential profiles presented in Figure \ref{fig:Figure4}b for both the horizontal and vertical cuts. A beam FWHM of 2.4 $\times$ 0.9\,\textmu m was determined from single Lorentzian fits. The pointing of the capillary was not adjusted after closing the scrapers for these measurements, however later realignment of the optic did not produce a reduced beam width. A minimum horizontal beam width of 1.3\,\textmu m was possible by selecting a different region of the capillary, however this could not be produced with a concurrent minimal vertical width. With these narrower beam profiles, particularly for the vertical direction, the peak approaches the expected Gaussian profile. The Lorentzian broadening may thus be ascribed to the summation of a number of small, slightly shifted peaks caused by slope errors in the optic. A practical utilization of this could be to design a custom beamstop at the entrance to the capillary which isolates only a small, well-focusing region of the optic, or to position a pinhole behind the capillary for similar effect \cite{Koch2018}.

\section{Spatially resolved ARPES}

\begin{figure*} [t!]
	\begin{center}
		\includegraphics[width=1\textwidth]{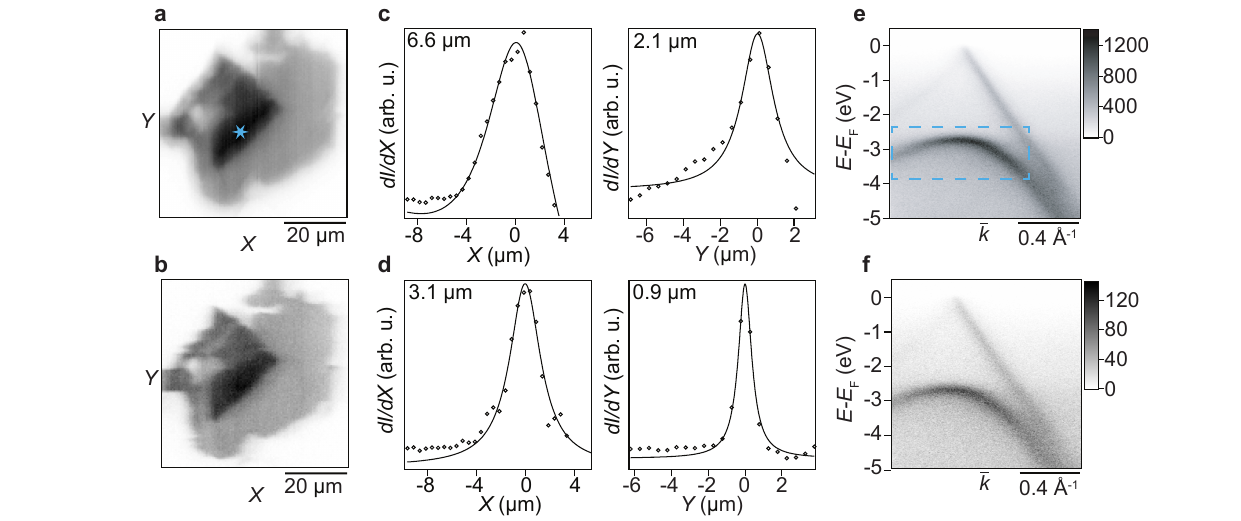}
		\caption{Photoemission measurements on a ML-hBN/Graphene/hBN heterostructure with a sub-micrometer beam. \textbf{a,} Photoemission map of a 2D graphene-hBN heterostructure measured with the optimally focussed capillary beam. \textbf{b,} Photoemission map measured with the apertured beam. The intensity from \textbf{a},\textbf{b} comes from the bulk hBN band. \textbf{c,} Horizontal (left) and vertical (right) differential intensity from a line measured across a sharp edge on the heterostructure, measured with the full capillary beam, fitted to a Lorentizian profile (solid line). \textbf{d,} Similar differential intensity profiles measured using the apertured beam. The FWHM of each Lorentzian is stated in the upper left. \textbf{e,} Photoemission spectra measured from the 2D heterostructure with the full capillary beam (top) and the sub-micrometer beam (bottom), measured from the position marked with a star in \textbf{a}. The dashed blue box denotes the region of interest providing intensity in \textbf{a},\textbf{b}.}
		\label{fig:Figure5}
	\end{center}
\end{figure*}

Confirming the use of a sub-micrometer beam, a 2D heterostructure device was measured using spatially resolved ARPES with the full and  apertured beam. This sample consisted of a heterostructure composed of bulk hBN underneath graphene underneath a monolayer (ML) of hBN, with lithographically patterned gold contacts electrically contacting the heterostructure at the sides. Spatially resolved ARPES maps were produced by raster scanning the beam across the sample and collecting a 2D $(E,k)$ ARPES image from each point. Such spatial maps are shown in Figures \ref{fig:Figure5}a and \ref{fig:Figure5}b for the full and apertured beam, respectively, where the intensity in the maps corresponds to the bulk hBN electronic structure within the blue dashed square in Figure \ref{fig:Figure5}e. The map in Figure \ref{fig:Figure5}b is clearly sharper than that in \ref{fig:Figure5}a, with narrow cracks becoming visible in the darker region of ML hBN. A slight blurring in the horizontal direction is apparent. Differential line profiles across sharp edges on the heterostructure are compared between the full beam in Figure \ref{fig:Figure5}c, and the apertured beam in Figure \ref{fig:Figure5}d. The beam spot is reduced from 6.6 $\times$ 2.1\,\textmu m to 3.1 $\times$ 0.9\,\textmu m by introducing the aperture. In the ARPES geometry, there is a 45$^\circ$ angle between the beam and the sample surface, which leads to horizontal broadening of the beam. Rough or angled edges on the sample may similarly increase the measured beam width both horizontally and vertically as the geometry  of the sample convolutes with  the profile of the beam. Previously in this work, we have assumed only the contribution from the beam while treating the sample or knife edge as perfectly sharp. These measurements confirm that the apertured beam is usable in a practical photoemission geometry to provide higher spatial contrast. Finally, photoemission spectra on the ML-hBN/ graphene region of the heterostructure were measured 
for the same duration using both the full capillary beam and the apertured beam in Figures \ref{fig:Figure5}e and \ref{fig:Figure5}f, respectively.
The bands are clearly reproduced by the apertured beam, with the measured photoemission intensity from the hBN band reduced by a factor of 10, providing an estimate of the reduction in transmission from this additional aperturing. This brings the spot size of this capillary optic towards a realisation of nanoARPES while retaining achromatic, high transmission.

\section{Conclusion}

The achromatic capillary optic is a powerful tool to produce a few-micrometer XUV beam spot, however it requires precise alignment, and thus stability, in order to provide the minimum spot size. The focal depth and variation with angle are determined for the capillary in use at AU-SGM4. The transmission was found to vary slowly across the full energy range of the AU-SGM4 beamline, confirming the optimal use of this optic on a synchrotron beamline. Slope errors, characterized by different beam profiles and focus behaviour for different regions of the capillary surface, lead to significant broadening of the beam. To overcome these limitations, sub-sections of the surface may be isolated to narrow the beam width, at the cost of lowered transmission. This could be further improved with custom-designed beam stops or using a separately controlled aperture before or ahead of the capillary to enable high spatial resolution measurements when required in an experiment.

\textbf{Data availability.} The data used in this study are available on Zenodo [\citenum{Jones2026}]

\section{Acknowledgements}
The work was funded/co-funded by the European Union (ERC grant EXCITE with project number 101124619). Views and opinions expressed are however those of the author(s)
only and do not necessarily reflect those of the European Union or the European Research
Council. Neither the European Union nor the granting authority can be held responsible for them. The authors acknowledge funding from the Novo Nordisk Foundation (Project Grants
NNF22OC0079960 and NFF23OC0085585) and the Danish Council for Independent Research (Grant No. 1026-00089B and 4258-00002B).

\section{Author information}
The authors declare that they have no competing financial interests.

\appendix
\setcounter{figure}{0}
\renewcommand{\thefigure}{A\arabic{figure}}
\section{Additional capillary characterization}
\label{app:capillary}

\begin{figure*} [t!]
	\begin{center}
		\includegraphics[width=1\textwidth]{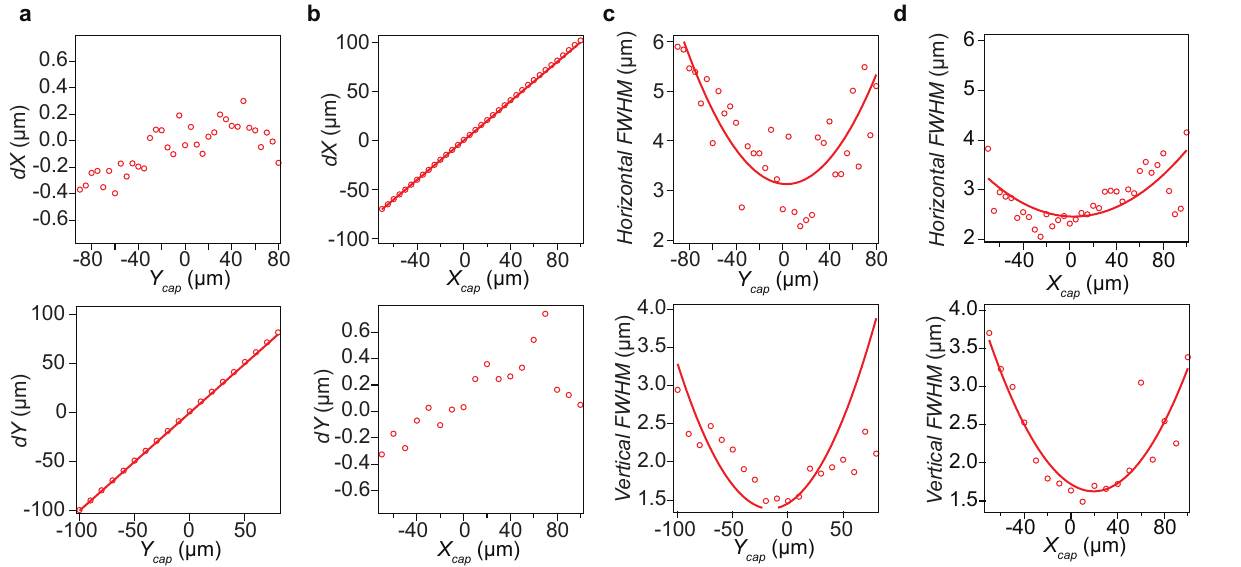}
		\caption{Focus behaviour with movement of capillary optic. \textbf{a,} Horizontal (top) and vertical (bottom) displacement of the beam position with vertical movement (\textit{Y}$_{cap}$) of the capillary. The solid line has a gradient of one. \textbf{b,} Displacement of the beam with horizontal movement of the capillary (\textit{X}$_{cap}$). \textbf{c,} Beam FWHM measured horizontally (top) and vertically (bottom) while moving the capillary vertically. A solid line marks a 2nd order polynomial fit as a guide to the eye. \textbf{d,} Beam FWHM measured while moving the capillary horizontally.}
		\label{fig:FigureA1}
	\end{center}
\end{figure*}

The capillary optic is intentionally over-illuminated by the synchrotron beam, allowing for perpendicular movement of the capillary with respect to the beam direction. This movement might be used for fine adjustment of the focus position or to modify the focus size. To determine the movement of the beam with motion of the capillary, the knife edge was translated through the focus and the resulting beam profile differentiated. The shift in the fitted peak of this differential profile horizontally (\textit{dX}) and vertically (\textit{dY}) is presented in Figure \ref{fig:FigureA1}a as a function of vertical motion of the capillary (Y$_{cap}$). The solid line in Figure \ref{fig:FigureA1}a depicts a gradient of one, demonstrating that the beam moves uniformly with the capillary position. The small horizontal variation of the beam with vertical position of the capillary in Figure \ref{fig:FigureA1}a may be explained by the knife edge not being set perpendicular to the beam. Figure \ref{fig:FigureA1}b shows the same measurement for horizontal motion of capillary (\textit{X}$_{cap}$), again demonstrating that the position of the beam shifts linearly with the optic's motion. However, this motion of the capillary was found to simultaneously broaden the beam. Figures \ref{fig:FigureA1}c and \ref{fig:FigureA1}d demonstrate a parabolic increase in the measured beam diameter both horizontally and vertically with the movement of the capillary optic in either \textit{Y} or \textit{X}. This may be explained by the short, 750\,mm distance between the intermediate focus and the capillary. The movement of the capillary changes the incident angle of the light, thus broadening the beam. 

\begin{figure*} [t!]
	\begin{center}
		\includegraphics[width=1\textwidth]{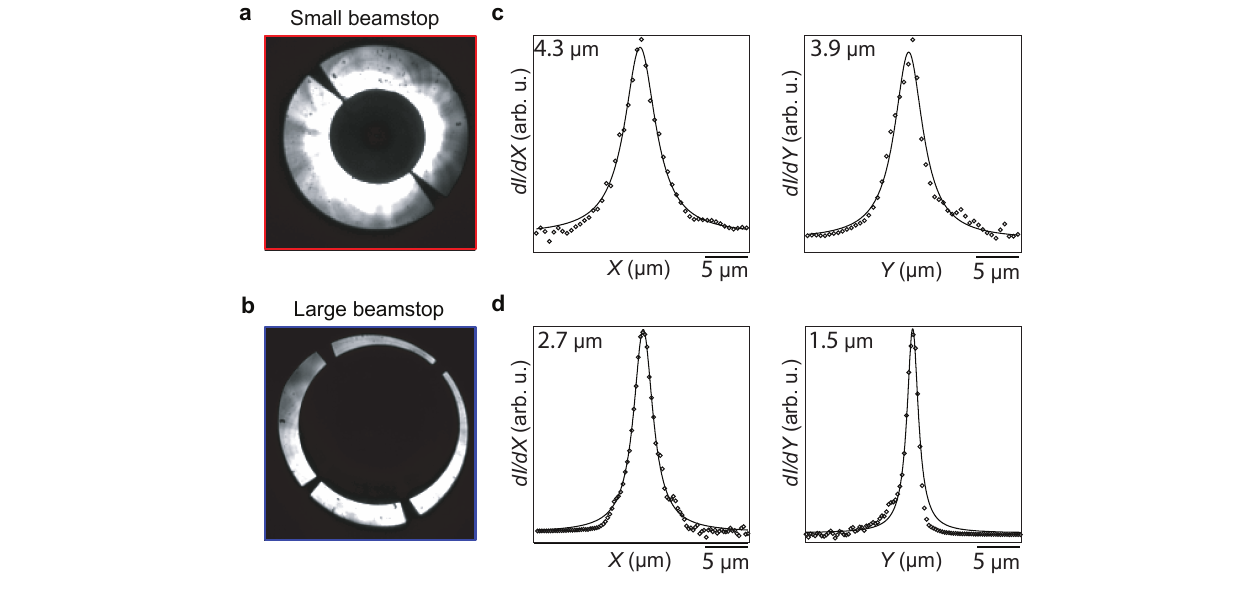}
		\caption{Comparison of different sized beamstops. \textbf{a,} Far field intensity of the capillary measured on the MCP with the small, minimally occluding beamstop. \textbf{b,} Far field intensity on the MCP with a larger beamstop overlaid on top of the smaller one. \textbf{c,} Optimal differential profiles (markers) and Lorentzian fits (solid lines) measured with the small beamstop horizontally (left) and vertically (right). \textbf{d,} Optimal differential profiles measured with the large beamstop. The Lorentzian FWHM is stated for each fit.}
		\label{fig:FigureA2}
	\end{center}
\end{figure*}

To determine the effect of different illumination of the capillary, two sizes of beamstop were used which occluded different amounts of the beam. Figure \ref{fig:FigureA2}a shows the far field image on the MCP screen with the smaller beamstop. Two arms supporting the central ring are visible. This original beamstop provides the minimum possible blocking of the beam, stopping only the light that would not reflect off of the capillary surface. A second, larger beamstop was then threaded on top of the original, greatly reducing the illumination of the optic, as shown in Figure \ref{fig:FigureA2}b. Two additional arms are visible due to misalignment between the arms on the two beamstops. The optimal focus was determined for each of the two beamstops, presented in Figures \ref{fig:FigureA2}c and \ref{fig:FigureA2}d for the small and large beamstops, respectively. The beam width is reduced from 4.3 $\times$ 3.9\,\textmu m to 2.7 $\times$ 1.5\,\textmu m at the cost of significant beam flux. This indicates that the inner and outer regions seen in the far field are focusing at different positions. The larger, more occluding beamstop is used for the measurements discussed in the main text, providing the smaller beam diameter.

%

\end{document}